\documentclass[apjl]{emulateapj}
\usepackage{apjfonts}
\usepackage{graphicx}

\shorttitle{Eccentric double white dwarfs in globular clusters}
\shortauthors{B. Willems et al.}

\begin{document}

\title{Eccentric double white dwarfs as LISA sources in globular clusters}

\author{B.\ Willems$^1$, V.\ Kalogera$^1$, A.\ Vecchio$^{1,2}$, N.\ Ivanova$^3$, F.\ A.\ Rasio$^1$, J.\ M.\ Fregeau$^1$, K.\ Belczynski$^{4,5}$}
\altaffiltext{1}{Northwestern University, Department of Physics and Astronomy, 2131
  Tech Drive, Evanston, IL 60208, USA}
\altaffiltext{2}{School of Physics and Astronomy, University of Birmingham, Edgbaston, Birmingham B15 2TT, UK}
\altaffiltext{3}{Canadian Institute for Theoretical Astrophysics, University of Toronto, 60 St. George, Toronto, ON M5S 3H8, Canada}
\altaffiltext{4}{New Mexico State University, Department of Astronomy, 1320 Frenger Mall, Las Cruces, NM 88003, USA}
\altaffiltext{5}{Tombaugh Fellow}

\begin{abstract}

We consider the formation of double white dwarfs (DWDs) through dynamical interactions in globular clusters. Such interactions can give rise to eccentric DWDs, in contrast to the exclusively circular population expected to form in the Galactic disk. We show that for a 5-year Laser Interferometer Space Antenna (LISA) mission and distances as far as the Large Magellanic Cloud, multiple harmonics from eccentric DWDs can be detected at a signal-to-noise ratio higher than 8 for at least a handful of eccentric DWDs, given their formation rate and typical lifetimes estimated from current cluster simulations. Consequently the association of eccentricity with stellar-mass LISA sources does not uniquely involve neutron stars, as is usually assumed. Due to the difficulty of detecting (eccentric) DWDs with present and planned electromagnetic observatories, LISA could provide unique dynamical identifications of these systems in globular clusters.
\end{abstract}


\section{Introduction}

Double white dwarf binaries (DWDs) are the single most abundant and a guaranteed class of gravitational wave (GW) sources for the {\it Laser Interferometer Space Antenna} (LISA; Bender et al.\ 1998). Their formation encompasses two mass-transfer phases, with at least one leading to a common-envelope phase shrinking the orbit to periods of 0.1--1000\,hr (Han 1998; Nelemans et al.\ 2001). Once formed, orbital angular momentum losses due to GW emission and tidal interactions continue to shrink the orbit until the least massive WD fills its Roche lobe and transfers mass to its companion. Depending on the binary mass ratio, the mass transfer leads to a fast merger or a long-lived phase of orbital expansion (Marsh et al. 2004; Gokhale et al. 2007, and references therein). 

As the progenitors of DWDs are tight enough to undergo multiple mass-transfer phases, tidal forces are usually thought to circularize their orbits long before they form a DWD. DWDs are therefore expected to be born with circular orbits. However, dynamical interactions in globular clusters (GCs) form DWDs at enhanced rates compared to the field (Shara \& Hurley 2002).  
Through these interactions, DWDs can form with non-zero eccentricities; these can subsequently be damped by GW emission, but can also be excited by fly-by encounters with single and binary stars. Contrary to common assumptions, the identification of eccentric stellar-mass binaries in the LISA GW data stream therefore does not uniquely identify them as neutron star (NS) binaries.

Due to their inherent faintness, DWDs are difficult to detect at GC distances or beyond, even with the SDSS, HST, or JWST in the future. LISA could therefore provide unique dynamical identifications of eccentric DWDs in GCs. As the potential of studying WD interiors through tidal effects increases dramatically with increasing orbital eccentricity, eccentric DWDs offer a unique opportunity to study stellar astrophysics with LISA.

In this Letter, we explore the formation of DWDs with non-zero eccentricities ($e>0.01$) and orbital periods $P < 5000$\,s through dynamical interactions in GCs, and assess their detectability by LISA. Since mass-transferring binaries are not expected to retain appreciable eccentricities for long periods of time, we focus our attention on detached DWDs. While the importance of stellar-mass binaries in GCs for LISA has been discussed previously (Benacquista 1999, Benacquista et al. 2001, Benacquista 2001), the possible presence of eccentric DWDs in the LISA band has not been considered before.

\section{Double white dwarfs in globular clusters}

Ivanova et al.\ (2006) recently explored the formation and evolution of WD binaries in GCs using an updated version of the Monte Carlo code described in Ivanova et al.\ (2005). In this study, we adopt these sample simulations to address the relevance of eccentric DWDs as LISA sources. The code combines a simple two-zone core-halo GC model with the {\sc startrack} binary population synthesis code (Belczynski et al.\ 2007) and the {\sc fewbody} small N-body integrator (Fregeau at al.\ 2005) to treat dynamical interactions between single stars and binaries. 

In the simulations, the two most important types of stellar interactions creating eccentric DWDs in the LISA band are exchange interactions and physical collisions. Exchange interactions are a subset of the broader class of binary-single and binary-binary interactions. 
Eccentric LISA DWDs form through this channel by exchanges between single WDs and binaries with a WD component. During the exchange, the two WDs become bound, while the companion to the initial binary WD is ejected from the system. The resulting binaries have average post-exchange eccentricities $e \simeq 0.8$. Alternatively, physical collisions between WDs and red giants form eccentric DWDs when the giant's envelope gets disrupted and its degenerate core becomes bound to the initially single WD (Lombardi et al. 2006). LISA DWDs formed this way have average post-collosion eccentricities $e \simeq 0.5$. We stress that the quoted average eccentricities are for LISA DWDs only (average post-exchange and post-collision eccentricities for the entire DWD population are lower).

In addition to direct formation, we find that non-zero eccentricities of DWDs can be significantly enhanced through one or multiple  binary-single star scattering events. These interactions are either distant fly-by encounters or close, strong, and possibly resonant interactions. Even for distant encounters (at more than 5--10 times the binary separation) and initial eccentricities $e<0.1$, Heggie \& Rasio (1996) find that the {\em relative} change in eccentricity can be as high as 30\%. Consequently, the {\em absolute} eccentricity change in a given interaction is insignificant for circular binaries, but increases with the binary's eccentricity and the cumulative effect of multiple interactions. On the other hand, close and possibly resonant interactions are rare, but can change the binary eccentricity drastically in just a single encounter.  

The possibility of forming eccentric DWDs through the above channels is in stark contrast to DWDs in the Galactic disk which are expected to be invariably circular. Despite the many dynamical interactions, circular binaries still dominate DWD populations in GCs, although we will show in the next section that the number of eccentric DWDs is non-ngligible. Circular DWD binaries could in principle also become eccentric via binary-single star encounters, although the efficiency of the process is small (Heggie \& Rasio 1996). Nevertheless, all possible eccentricity inducing effects are included in the simulations described below, regardless of their relative efficiency. Since the detectability and properties of GW signals from eccentric DWDs are independent of the details of their formation, we do not distinguish between the different formation channels in the remainder of this paper.

\section{Monte Carlo simulations}

We use simulations by Ivanova et al. (2006) to estimate the expected number of eccentric DWDs in the LISA band ($10^{-4}$--$10^{-1}$\,Hz). The outcome of the simulations depends on the adopted GC model parameters (e.g., cluster core density, metallicity) and stellar and binary evolution parameterizations (e.g., initial masses and orbital elements, efficiency of orbital angular momentum loss mechanisms). We consider three GC models representative of a typical Galactic, 47 Tuc-like, and Terzan 5-like GC (see Table~\ref{models}), and GC ages in the range of 10--13\,Gyr. All stellar and binary evolutionary phases are treated as in the standard model of Ivanova et al.\ (2006). 

\begin{deluxetable}{lccccc}
\tablecolumns{6}
\tablecaption{Globular cluster model parameters. \label{models}}
\tablehead{
   \colhead{GC model} & \colhead{$n_{\rm c}$} & \colhead{$t_{\rm rh}$} & \colhead{$Z$} & \colhead{$\sigma_1$} & \colhead{$v_{\rm esc}$} 
   }
\startdata
Typical  & $10^5$            & 1.0  & 0.005  & 10   & 40    \\
47 Tuc-like   & $2.5 \times 10^5$ & 3.0  & 0.0035 & 11.5 & 57   \\
Terzan 5-like & $8 \times 10^5$   & 0.93 & 0.013  & 11.6 & 49 
\enddata
\tablecomments{GC core number density $n_{\rm c}$ (${\rm pc}^{-3}$), half-mass relaxation time $t_{\rm rh}$ (Gyr), metallicity $Z$, one-dimensional velocity dispersion $\sigma_1$ (${\rm km\,s^{-1}}$), and escape velocity $v_{\rm esc}$ (${\rm km\,s^{-1}}$).}
\end{deluxetable}

The maximum number of eccentric DWDs present in the LISA band at any given time in the 10--13\,Gyr time interval is 2, 8, and 7, for the typical, 47 Tuc-like, and Terzan 5-like GC models, respectively. Using the total number of systems formed in the 10--13\,Gyr time window, we calculate the formation rates of eccentric LISA DWDs in the typical, 47 Tuc-like, and Terzan 5-like GC models to be $\simeq10^{-8}\,{\rm yr^{-1}}$, $\simeq3 \times 10^{-8}\,{\rm yr^{-1}}$, and $\simeq5 \times 10^{-8}\,{\rm yr^{-1}}$.  For each GC model, the probability that $N$ eccentric LISA DWDs are present at any given time in the 10--13\,Gyr age range is furthermore well fitted by a Poisson distribution with mean $ \langle N \rangle$ listed in Table~\ref{Poisson}. It follows that the probability to find at least 1 eccentric DWD in the LISA band is 32\% for the typical Galactic, 84\% for the 47 Tuc-like, and 80\% for the Terzan 5-like GC model. 

\begin{figure*}
\begin{center}
\resizebox{17.0cm}{!}{\includegraphics{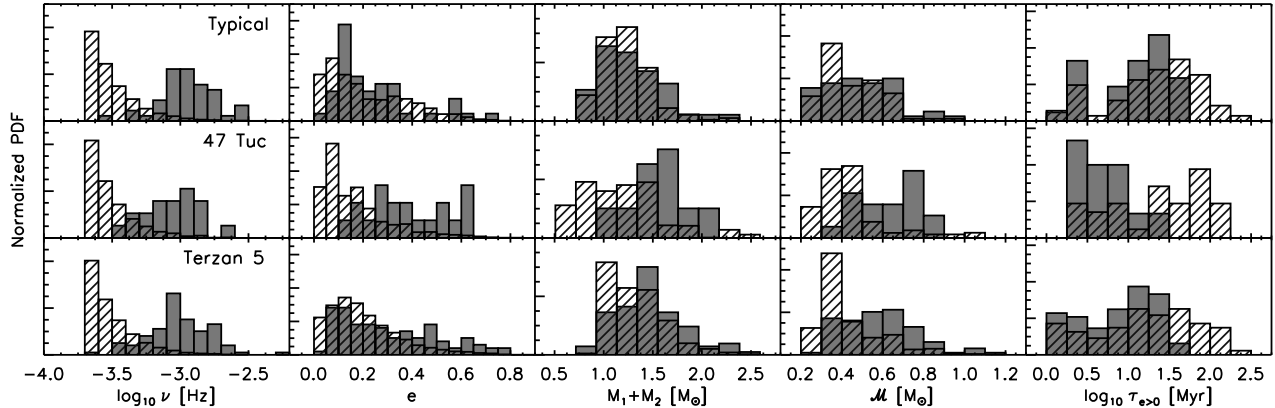}}
\caption{Statistical properties of eccentric DWDs ($e > 0.01$) in the LISA band ($P < 5000$\,s). Hatched histograms represent the entire eccentric DWD population; grey shaded histograms represent the subpopulation of systems for which at least 2 harmonics are individually observable by LISA at ${\rm SNR} > 8$, assuming a 5 year LISA mission, a distance of 10\,kpc, and a single time-delay interferometry observable. As the PDFs are normalized to unity, the vertical scale varies from panel to panel.}
\label{hist}
\end{center}
\end{figure*} 

The hatched probability distribution functions (PDFs) in Fig.~\ref{hist} show the statistical properties of the eccentric LISA DWDs in the three GC simulations. For all three models, the number of systems decreases monotonically with increasing orbital frequency $\nu = 1/P$. The abrupt drop in the number of systems at $\log \nu \la -3.7$ corresponds to the adopted cut-off orbital period of 5000\,s for sources in the LISA band. The orbital eccentricity distributions (affected by post-formation GW emission) typically peak at $e \simeq 0.1$ and decrease slowly toward higher values. In all three cases, there is a non-negligible fraction of systems with $e$ up to $\simeq 0.7$. The majority of systems consists of at least one WD more massive than $\sim 0.6\,M_\odot$.  The total mass distributions therefore peak at relatively high masses $M_1+M_2 \simeq 1.0$--$1.2\,M_\odot$. The chirp mass distributions peak at ${\cal M} \simeq 0.3$--$0.4\,M_\odot$. The typical merger lifetime $\tau_{e > 0}$ of an eccentric LISA DWD is 10--100\,Myr. However, a substantial number of systems also have considerably shorter lifetimes down to $\sim 1$\,Myr.

\begin{deluxetable}{lcccc}
\tablecolumns{5}
\tablecaption{Statistical properties of eccentric LISA DWDs. \label{Poisson}}
\tablehead{
   \colhead{GC model} & \colhead{$\langle N \rangle$} & \colhead{$P_1$} & \colhead{$P_2$} & \colhead{$P_3$}
   }
\startdata
Typical  & 0.39 (0.01) & 0.32 (0.01) & 0.06 & 0.01 \\
47 Tuc-like   & 1.9  (0.03) & 0.84 (0.03) & 0.55 & 0.29 \\
Terzan 5-like & 1.6  (0.03) & 0.80 (0.03) & 0.48 & 0.22
\enddata
\tablecomments{Mean number $\langle N\rangle$ of eccentric LISA DWDs and Poisson probabilities $P_N$ that at least $N$ such DWDs are present at any given time. Numbers in parentheses correspond to DWDs with at least two harmonics detectable at a distance of 10\,kpc with ${\rm SNR} > 8$, for a 5 year LISA mission and a single time-delay interferometry observable.}
\end{deluxetable}

\section{Implications for LISA}

Gravitational waves from binaries with non-zero eccentricity produce radiation at frequencies that are integer multiples of the orbital frequency. To compute the LISA sensitivity to eccentric DWDs, we model GWs at the leading quadrupole order. The (angle-averaged) SNR at which a given binary can be detected over an observation time $T_\mathrm{obs}$ is then given by
\begin{equation}
\left\langle \left(\mathrm{S}/\mathrm{N}\right)^2 \right\rangle = 
\sum_n \left\langle \left(\mathrm{S}/\mathrm{N}\right)_n^2 \right\rangle = 
\sum_n \int \! \left[\frac{h_{\mathrm{c},n}(f_n)}{h_\mathrm{rms}(f_n)}\right]^2 d\ln f_n,
\label{e:snr}
\end{equation}
where $n$ labels the harmonics at frequency $f_n \simeq n\,\nu$, 
\begin{equation}
h_{\mathrm{c},n} = (\pi d)^{-1}\,\left( 2\, \dot{E}_n/\dot{f}_n \right)^{1/2}
\label{e:hc}
\end{equation}
is the characteristic amplitude of the $n$-th harmonic, and $h_\mathrm{rms}^2(f_n) = (20/3)\,S_h(f_n)\,f_n$ is the root-mean square value of the noise averaged over source position and orientation; $S_h(f_n)$ is the one-sided noise spectral density, which we computed using Larson's  online sensitivity curve generator\footnote{http://www.srl.caltech.edu/$\sim$shane/sensitivity/}. In Eq.~(\ref{e:hc}), $\dot{E}_n$ is the time derivative of the energy radiated in GWs at frequency $f_n \simeq n\nu$, which to lowest order is
\begin{equation}
\dot{E}_n = (32/5)\left(2 \pi \nu {\cal M}\right)^{10/3} g(n,e).
\label{e:dotEn}
\end{equation}
Here, $g(n,e)$ is a function of the orbital eccentricity given by Eq. (20) in Peters \& Mathews (1963). The term $\dot{f}_n \simeq n\, d{\nu}/dt$, to the leading quadrupole order, is 
\begin{equation}
\dot{f}_n = 96\,(10\pi)^{-1}\,{\cal M}^{5/3}(2 \pi \nu)^{11/3}\,F(e)\,,
\label{e:dotnu}
\end{equation}
where $F(e) = [{1 + (73/24)\,e^2 + (37/96)\,e^4}]/{(1 - e^2)^{7/2}}$. The latter function tends to unity for $e\rightarrow 0$, but can become of the order of a few for $e$-values typical of the systems produced in our simulations, {\em e.g.} $F(0.2) = 1.29$ and $F(0.5) = 4.88$. 

Using Eqs.~(\ref{e:snr})-(\ref{e:dotnu}), we compute the number of {\em detectable} eccentric LISA sources, requiring that at least two harmonics be {\em individually} observable at SNR of 8 or higher. This threshold is conservative, since we still do not exactly know how the data analysis will be carried out (note that in the LISA literature a ${\rm SNR}  = 5$ threshold is usually considered). The choice of assuming that at least two harmonics are observable separately, is dictated by reasons of robustness: from an analysis point of view, an eccentric DWD looks like a collection of $n$ "circular" (one single harmonic) binaries, with suitable amplitudes and frequencies. There is growing evidence that data analysis schemes that are already in hand for circular binaries (cf. {\em e.g.} Arnaud et al. 2007, and references therein) will perform well at such moderate-to-high SNR. We can therefore be confident that even an analysis, comprising of a first pass of the data to search for "circular" binaries, followed by a second stage in which one looks for a more consistent fit to the data by allowing that some signals are actually harmonics of the same system is available. In practice, it is quite conceivable that a full-blown analysis allowing for non-negligible eccentricity can be carried out from the beginning. In this case the number of detectable systems per GC presented here is conservative and should be regarded as a lower limit. 

\begin{figure}
\resizebox{8.0cm}{!}{\includegraphics{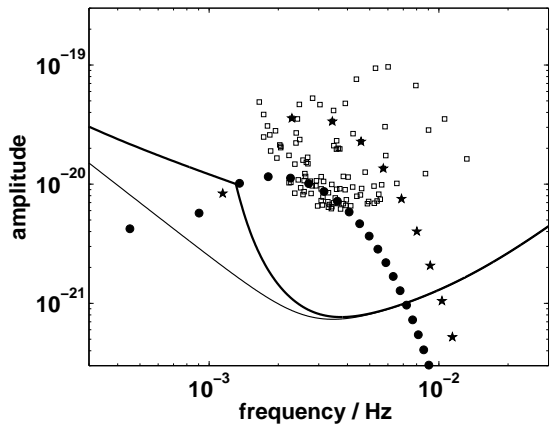}}
\caption{LISA sensitivity to eccentric DWDs in the typical Galactic GC model. The solid lines show the angle-averaged level of the rms noise (thin: instrumental noise; thick: total instrumental and galactic foreground noise). Squares indicate the amplitudes of DWDs for which at least two harmonics can be detected individually at ${\it SNR} > 8$.  Some squares correspond to the same system at different times in its evolution. For illustration, the solid stars and circles represent the harmonics of a DWD with masses $M_1=0.74\,M_\odot$ and $M_2=0.86\,M_\odot$ at two different stages of its evolution: at $P = 872.64$\,s and $e = 0.33$ (stars) and $P = 2211.84$\,s and $e = 0.74$ (circles). In both cases 4 harmonics are observable individually at ${\rm SNR} > 8$, but all $n = 1,\dots, 20$ harmonics are plotted for illustration. The amplitudes are for a 5-year mission, a GC distance of 10\,kpc, and a single Michelson observable.}
\label{f:sens}
\end{figure}

In Fig.~\ref{f:sens}, we show the detectability results for the typical Galactic GC model, adopting a distance of 10\,kpc. The amplitude on the vertical axis is the characteristic amplitude $h_{c,n}$ multiplied by the square root of $\min[1,df_n/dt\times (T_\mathrm{obs}/f_n)]$ to consistently account for the frequency band swept by each source (all the systems in our simulations satisfy the condition $df_n/dt\times (T_\mathrm{obs}/f_n) < 1$ for $T_\mathrm{obs} = 5$ yr). The height of each dot above the noise level then represents the relevant SNR value  [cf. Eq.~(\ref{e:snr})]. We consider $T_\mathrm{obs} = 5$ yrs, the current requirement for the minimum mission lifetime. We also use only one Michelson observable in the analysis; if all six Doppler links are available over $T_\mathrm{obs}$, the SNR would increase by a factor $\simeq \sqrt{2}$. Note that only systems for which at least two harmonics have $\langle (\mathrm{S}/\mathrm{N})_n^2 \rangle^{1/2} > 8$ are plotted in the figure. 

\begin{deluxetable}{lcccc}
\tablecolumns{5}
\tablecaption{Observable binaries. \label{t:nsystems}}
\tablehead{
   \colhead{GC model} & $N_{\rm tot}$ & $N_{\rm obs}$ at $3\,{\rm kpc}$ & $N_{\rm obs}$ at $10\,{\rm kpc}$ & $N_{\rm obs}$ at $50\,{\rm kpc}$
    }
\startdata
Typical  & 29   & 24 (28)  & 17 (27)  & 1 (7)   \\
47 Tuc-like   & 102   & 72 (96)  & 36 (90)  & 9 (24) \\
Terzan 5-like & 140 & 112 (133) & 79 (127) & 4 (36) 
\enddata
\tablecomments{Total number of eccentric LISA DWDs $N_{\rm tot}$ and number of systems $N_{\rm obs}$ with at least two harmonics \textrm{individually} detectable by LISA at distances of 3, 10, and 50\,kpc with a coherent ${\rm SNR} > 8$ for a 5 year LISA mission. Both $N_{\rm tot}$ and $N_{\rm obs}$ are integrated over a 10--13\,Gyr age range. Note that essentially all systems in GCs at Galactic distances are detectable with at least one harmonic; this number is provided within parenthesis in the $N_{\rm obs}$ columns. }
\end{deluxetable}

In Table~\ref{t:nsystems}, we list the number of detectable eccentric DWDs formed in the 10--13\,Gyr age range, for distances $d=3, 10, 50$\,kpc. Depending on the GC model, the numbers range from 24--112 for $d=3$\,kpc and 17--79 for $d=10$\,kpc. It is quite remarkable that LISA has a high enough sensitivity to observe these systems throughout the Galaxy and, for a very limited region of the parameter space, even up to the Large Magellanic Cloud. The probability to find at least 1 detectable eccentric DWD is 1\% for the typical Galactic GC model and 3\% for both the 47 Tuc-like and Terzan 5-like GC models (see Table~\ref{Poisson}). If we require only one harmonic to be detectable, the probabilities increase by a factor of $\sim 2$. Compared to the full population, the detectable samples of eccentric DWDs furthermore favor higher orbital frequencies, total system masses, and chirp masses, and shorter system lifetimes; relevant orbital eccentricities span the entire range from $e \simeq 0.1$ to 0.8 (grey PDFs in Fig.~\ref{hist}).

\section{Discussion}

In this Letter, we examine the formation of eccentric DWDs in the LISA band through dynamical interactions in GCs. Although the majority of cluster DWDs are circular, the existence of this non-negligible population is in stark contrast to the Galactic disk population expected to consist exclusively of circular binaries. Owing to the non-zero orbital eccentricities, the DWDs emit GWs at multiple harmonics of the orbital frequency. For a 5-year LISA mission, we find a non-negligible number of systems for which multiple harmonics are detectable individually at ${\rm SNR} > 8$, allowing the determination of the orbital eccentricity from the ratio of the amplitudes of the detected harmonics.

As a proof of principle, we show that for standard stellar and binary evolution assumptions, 47 Tuc and Terzan 5 type clusters are the most efficient eccentric DWD factories. For these two types of GCs, there is an 80\% probability that at least one eccentric DWD is present in the LISA band at any given time. For a typical Galactic GC, this probability is 32\%. Moreover, for a GC at 10\,kpc, the probability that one or more systems are present with at least two harmonics individually detectable at ${\rm SNR} > 8$ is 1\% for a typical Galactic GC, and 3\% for 47 Tuc and Terzan 5 type GCs. Considering there are 79 known GCs within 10\,kpc from the Sun (Harris 1996), we thus expect at least a handful of eccentric DWDs to be detectable by LISA. There are furthermore 3 GCs within 2--3\,kpc and 15 GCs within 5\,kpc from the Sun. For these clusters, the SNR increases by a factor of $\sim 2$--3. These numbers may evidently vary significantly with the adopted stellar and binary evolution model parameters. Since our goal was to assess the feasibility of forming and detecting eccentric DWDs in GCs, we chose parameters which are common in standard population synthesis models.

In principle, the independent knowledge of GC distances from electromagnetic observations allows the derivation of chirp masses of GC binaries from the amplitude of their GW signals. Unfortunately, the identification of sources with GCs is severely hampered by LISA's limited angular resolution. In particular, the error box in the sky is $\simeq 1.6\, (5\,\mathrm{mHz}/\nu)^2\, (10/\mathrm{SNR})^2$\,deg$^2$ (e.g. Takahashi \& Seto 2002) compared to a typical GC's size of $< 0.002$\,deg$^2$.  Independent chirp mass and source distance determinations therefore rely completely on measurements of intrinsic changes of the GW frequency during the observation time (e.g. Takahashi \& Seto 2002), assuming the drifts are entirely due to gravitational radiation with no significant contribution from tidal and/or magnetic spin-orbit coupling (e.g. Stroeer et al. 2005). It is therefore interesting to note that, depending on their physical parameters and the phase of evolution at which they are observed, the systems in our simulations can produce observable frequency drifts. 

We conclude that the identification of eccentric binaries with LISA does not necessary imply association with NSs. Since binary component masses are difficult to derive from GW data, a clear discrimination between NS-WD or NS-NS and DWD binaries from LISA data may not be possible. However, the number of eccentric NS-WD binaries in the three GC models is an order of magnitude lower than the number of eccentric DWDs, and NS-NS binaries are even scarcer. Statistically, an eccentricity detection therefore favors a DWD interpretation over a NS-WD or NS-NS interpretation. The formation of eccentric DWDs in GCs thus provides an exciting opportunity to study WD physics inaccessible through DWDs in the Galactic disk. Tidal effects particularly become increasingly important with increasing orbital eccentricity, either through the excitation of dynamical oscillations or through their effects on the orbital evolution. LISA  could provide unique dynamical identifications of eccentric DWDs, opening up a unique astrophysical window to stellar and binary evolution.

\acknowledgments 

This work is partially supported by a Packard Foundation Fellowship, a NASA BEFS grant (NNG06GH87G), and a NSF CAREER grant (AST-0449558) to VK. FAR and JMF acknowledge support from NASA grants NNG06GI62G and NNG04G176G, and from NSF grant PHY-0601995. Cluster simulations were performed on CITA's Sunnyvale cluster, funded by the CFI and the ORF-RI.


\begin{thebibliography}{}
\bibitem[Arnaud et al.(2007)]{2007gr.qc.....1139A} Arnaud, K.~A., et al.\ 
2007, arXiv:gr-qc/0701139
\bibitem[Belczynski et al.(2007)]{2005astro.ph.11811B} Belczynski, K., et al.\ 2007, in press, arXiv:astro-ph/0511811
\bibitem[Bender et al.(1998)], Bender,~P. et al.\ 1998 LISA Pre-Phase A Report, 2nd Ed. (MPQ)
\bibitem[Benacquista(1999)]{1999ApJ...520..233B} Benacquista, M.\ 1999, 
\apj, 520, 233
\bibitem[Benacquista(2001)]{2001AIPC..586..793B} Benacquista, M.\ 2001, 
20th Texas Symposium on relativistic astrophysics, 586, 793
\bibitem[Benacquista et al.(2001)]{2001CQGra..18.4025B} Benacquista, M.~J., 
Portegies Zwart, S., \& Rasio, F.~A.\ 2001, Classical and Quantum Gravity, 
18, 4025
\bibitem[Fregeau et al.(2004)]{2004MNRAS.352....1F} Fregeau, J.~M., Cheung, P., Portegies Zwart, S.~F., \& Rasio, F.~A.\ 2004, \mnras, 352, 1 
\bibitem[Gokhale et al.(2007)]{2007ApJ...655.1010G} Gokhale, V., Peng, 
X.~M., \& Frank, J.\ 2007, \apj, 655, 1010 
\bibitem[Han(1998)]{1998MNRAS.296.1019H} Han, Z.\ 1998, \mnras, 296, 1019
\bibitem[Harris(1996)]{1996AJ....112.1487H} Harris, W.~E.\ 1996, \aj, 112, 
1487 
\bibitem[Heggie \& Rasio(1996)]{1996MNRAS.282.1064H} Heggie, D.~C., \& 
Rasio, F.~A.\ 1996, \mnras, 282, 1064
\bibitem[Ivanova et al.(2005)]{2005MNRAS.358..572I} Ivanova, N., Belczynski, K., Fregeau, J.~M., \& Rasio, F.~A.\ 2005, \mnras, 358, 572
\bibitem[Ivanova et al.(2006)]{2006MNRAS.372.1043I} Ivanova, N., Heinke, C.~O., Rasio, F.~A., Taam, R.~E., Belczynski, K., \& Fregeau, J.~A.\ 2006, \mnras, 372, 1043
\bibitem[Larson(2002)]{online-sensitivity} Larson, S.~L. 2007, Online LISA sensitivity curve generator 
\bibitem[Lombardi et al.(2006)]{2006ApJ...640..441L} Lombardi, J.~C., Jr., Proulx, Z.~F., Dooley, K.~L., Theriault, E.~M., Ivanova, N., \& Rasio, F.~A.\ 2006, \apj, 640, 441
\bibitem[Marsh et al.(2004)]{2004MNRAS.350..113M} Marsh, T.~R., Nelemans, 
G., \& Steeghs, D.\ 2004, \mnras, 350, 113
\bibitem[Nelemans et al.(2001)]{2001A&A...365..491N} Nelemans, G., Yungelson, L.~R., Portegies Zwart, S.~F., \& Verbunt, F.\ 2001, \aap, 365, 491
\bibitem[Peters \& Mathews(1963)]{PM1963} Peters, P. C., \& Mathews, J. 1963, Phys. Rev., 131, 435
\bibitem[Peters(1964)]{P1964} Peters, P. C. 1964, Phys. Rev., 136, 1224
\bibitem[Shara \& Hurley(2002)]{2002ApJ...571..830S} Shara, M.~M., \& 
Hurley, J.~R.\ 2002, \apj, 571, 830
\bibitem[Stroeer et al(2005)]{Stroeer05} Stroeer, A., Vecchio, A. \& Nelemans, G. \ 2005, \apj, 633, L33
\bibitem[Takahashi \& Seto(2002)]{2002ApJ...575.1030T} Takahashi, R., \& 
Seto, N.\ 2002, \apj, 575, 1030
\end{thebibliography}
\end{document}